\documentclass{PoS}

\newcommand{\be}{\begin{equation}}
\newcommand{\ee}{\end{equation}}
\def\bea{\begin{eqnarray}}
\def\eea{\end{eqnarray}}
\def\bean{\begin{eqnarray*}}
\def\eean{\end{eqnarray*}}

\title{
{\small \normalfont \vspace*{-6.0cm} \hspace*{11.0cm} CERN-PH-TH/2006-215 \vspace*{5.0cm}} \\
Towards the QCD phase diagram 
}

\ShortTitle{Towards the QCD phase diagram}

\author{\speaker{Philippe de Forcrand}\\
        Institut f\"ur Theoretische Physik, ETH Z\"urich, CH-8093 Z\"urich, Switzerland ~ and \\
        CERN, Physics Department, TH Unit, CH-1211 Geneva 23, Switzerland \\
        E-mail: \email{forcrand@phys.ethz.ch}}
\author{Owe Philipsen\\
        Institut f\"ur Theoretische Physik, Westf\"alische Wilhelms-Universit\"at M\"unster, Germany \\
        E-mail: \email{ophil@uni-muenster.de}}

\abstract{
We summarize our recent results on the phase diagram of QCD with $N_f=2+1$ quark flavors,
as a function of temperature $T$ and quark chemical potential $\mu$.
Using staggered fermions, lattices with temporal extent $N_t=4$,
and the exact RHMC algorithm, we first determine the critical line in the quark mass plane 
$(m_{u,d},m_s)$ where the finite temperature transition at $\mu=0$ is second order.
We confirm that the physical point lies on the crossover side of this line.
Our data are consistent with a tricritical point at $(m_{u,d},m_s) = (0,\sim 500)$ MeV.

Then, using an imaginary chemical potential, we determine in which direction
this second-order line moves as the chemical potential is turned on.
Contrary to standard expectations, we find that the region of first-order transitions
shrinks in the presence of a chemical potential,
which is inconsistent with the presence of a QCD critical point at small chemical potential.

The emphasis is put on clarifying the translation of our results from lattice
to physical units, and on discussing the apparent contradiction of our findings
with earlier lattice studies.
}

\FullConference{XXIVth International Symposium on Lattice Field Theory\\
                July 23-28, 2006\\
                Tucson, Arizona, USA}

\begin{document}

\begin{figure}[t!]
\centerline{
\scalebox{0.40}{\includegraphics{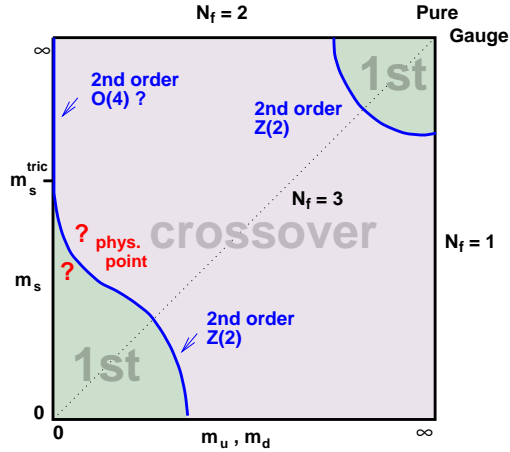}}
}
\caption{Schematic phase transition behaviour of $N_f=2+1$ flavor QCD for
different choices of quark masses $(m_{u,d},m_s)$, at $\mu=0$ (from \cite{OP_review}).}
\end{figure}

\section{Introduction}

In recent years, considerable efforts have been devoted to the determination of
the phase diagram of QCD at finite temperature and density~\cite{OP_review}. 
At zero chemical potential, the behaviour as a function of temperature depends
on the quark masses $m_{u,d}$ and $m_s$, and expectations are summarized in Fig.~1.
In the limits of zero and infinite quark masses (lower left and upper right corners),
order parameters corresponding to the breaking of a symmetry can be defined, 
and one finds numerically that a first-order transition takes place at a finite
temperature $T_c$. On the other hand, one observes an analytic crossover at 
intermediate quark masses. Hence, each corner must be surrounded by a region of
first-order transition, bounded by a second-order line as in Fig.~1. 
The line in the heavy-quark corner has been studied in \cite{heavy}.
Here, we want to determine the chiral critical line.

Along both lines, the universality class is that of the $3d$ Ising model. 
Therefore, a powerful tool to determine the critical couplings 
is the Binder cumulant $B_4 \equiv \langle \delta X^4 \rangle / \langle \delta X^2 \rangle^2$,
where $\delta X = X - \langle X \rangle$, and we take for $X$ the $u,d$ quark condensate
$\bar\psi \psi$. On lattices $8^3, 12^3$ and $16^3 \times 4$,
we estimate the critical couplings as those for which $B_4 = 1.604..$, the $3d$ Ising
value. For each mass point $(m_{u,d},m_s)$, we accumulate at least 200k 
RHMC trajectories, and interpolate among 4 or more $m_{u,d}$ values to find the 
critical $m_{u,d}$ mass $m^c$ at a given $m_s$. 
We obtain the set of points in Fig.~4, left.

We then consider the effect of a baryonic chemical potential.
As a function of $\mu$, represented vertically in Fig.~2, the critical line 
determined at $\mu=0$ spans a surface. The standard expectation is depicted in 
Fig.~2 left. The first order region expands as $\mu$ is turned on, so that the
physical point, initially in the crossover region, eventually belongs to the
critical surface. At that chemical potential $\mu_E$, the transition is second order:
that is the QCD critical point. Increasing $\mu$ further makes the transition
first order. A completely different scenario arises if instead the first-order
region shrinks as $\mu$ is turned on. In that case (Fig.~2 right), the physical
point remains in the crossover region for any $\mu$.
Since the phenomenologically interesting question is whether a QCD critical
point $(\mu_E,T_E)$ exists at small $\mu_E, \mu_E/T_E \lesssim 1$, this question
can be addressed by an analytic continuation based on a Taylor expansion. Using an imaginary chemical
potential~\cite{OP_Nf2,DElia}, we determine the curvature
$\frac{d m^c}{d \mu^2}|_{\mu=0}$ of the critical surface at $\mu=0$.
We find that it is negative, so that the first-order region shrinks as in
Fig.~2 right. Note that in the opposite corner, the first-order region also
shrinks~\cite{Potts}.

Sec.~2 tests our methodology in the $N_f=3$ case. Sec.~3 describes the $N_f=2+1$
study. Sec.~4 compares our results with earlier lattice studies and discusses the
various limitations of our approach.

\begin{figure}[t!]
\centerline{
\scalebox{0.65}{\includegraphics{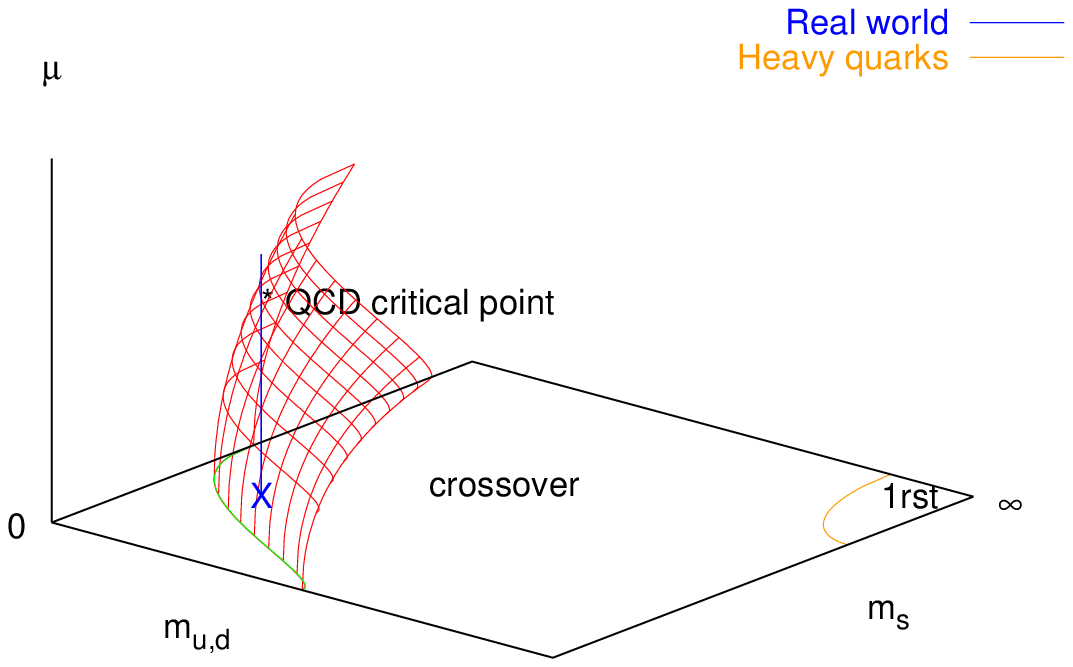}}
\scalebox{0.65}{\includegraphics{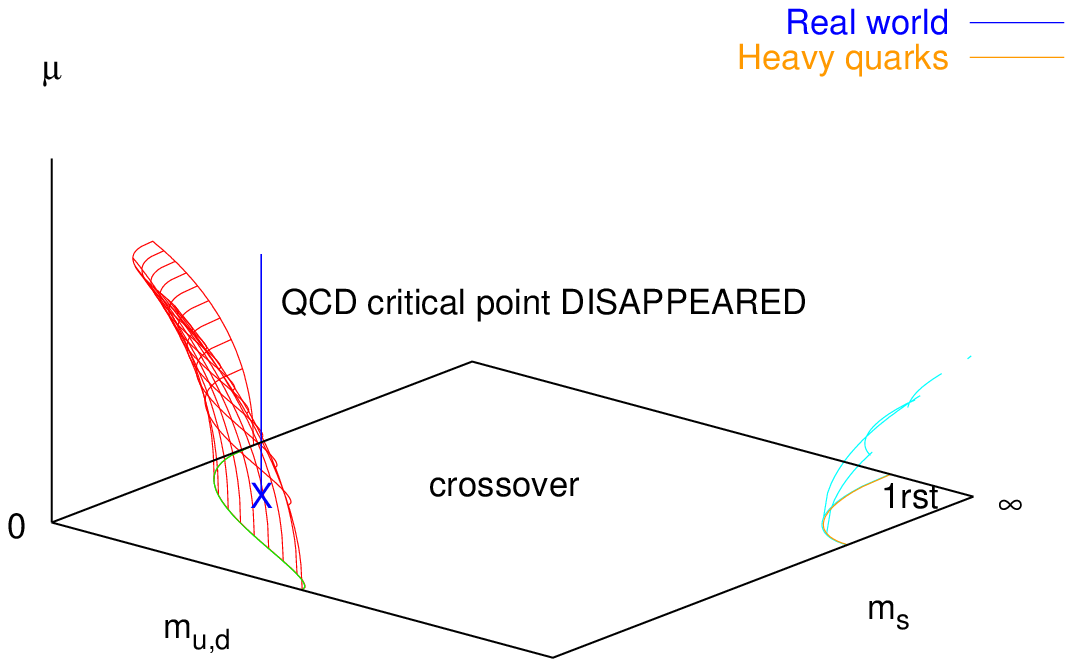}}
}
\vspace*{-0.5cm}
\caption{The chiral critical surface in the case of positive (left) and negative (right) curvature.
If the physical point is in the crossover region for $\mu=0$, a finite $\mu$
phase transition will only arise in the scenario (left) with positive curvature,
where the first-order region expands with $|\mu|$.
Note that for heavy quarks, the first-order region shrinks with $|\mu|$ (right)
~\cite{Potts}.}
\end{figure}

\section{$N_f=3$}
\label{sec_Nf3}

We first check our methodology in the case of 3 degenerate flavors.
This is basically a repeat of Ref.~\cite{OP_Nf3}, this time using the
RHMC algorithm~\cite{RHMC} instead of the R algorithm.

The RHMC algorithm eliminates the stepsize error of the R algorithm,
which differs in magnitude in the chirally symmetric and broken phases
\cite{DKS_R}. As a result, the value of $m^c(\mu=0)$ is considerably
different: $(a m^c(\mu=0))$ moves from 0.033(1) (R alg.)~\cite{OP_Nf3}
to 0.0260(5) (see Fig.~3, left). We have checked, by performing zero-temperature
simulations at this quark mass, that this is not a simple renormalization
effect, but that the physical ratio $m_\pi/T_c$ is lowered by about 10\%.
Therefore, an exact algorithm appears mandatory for the study of the
$N_f=2+1$ critical line. Moreover, RHMC turns out to be vastly more
efficient, by up to a factor 20 in our case for the smallest quark
masses~\cite{RHMC_LAT05}.

\begin{figure}[t!]
\centerline{
\scalebox{0.62}{\includegraphics{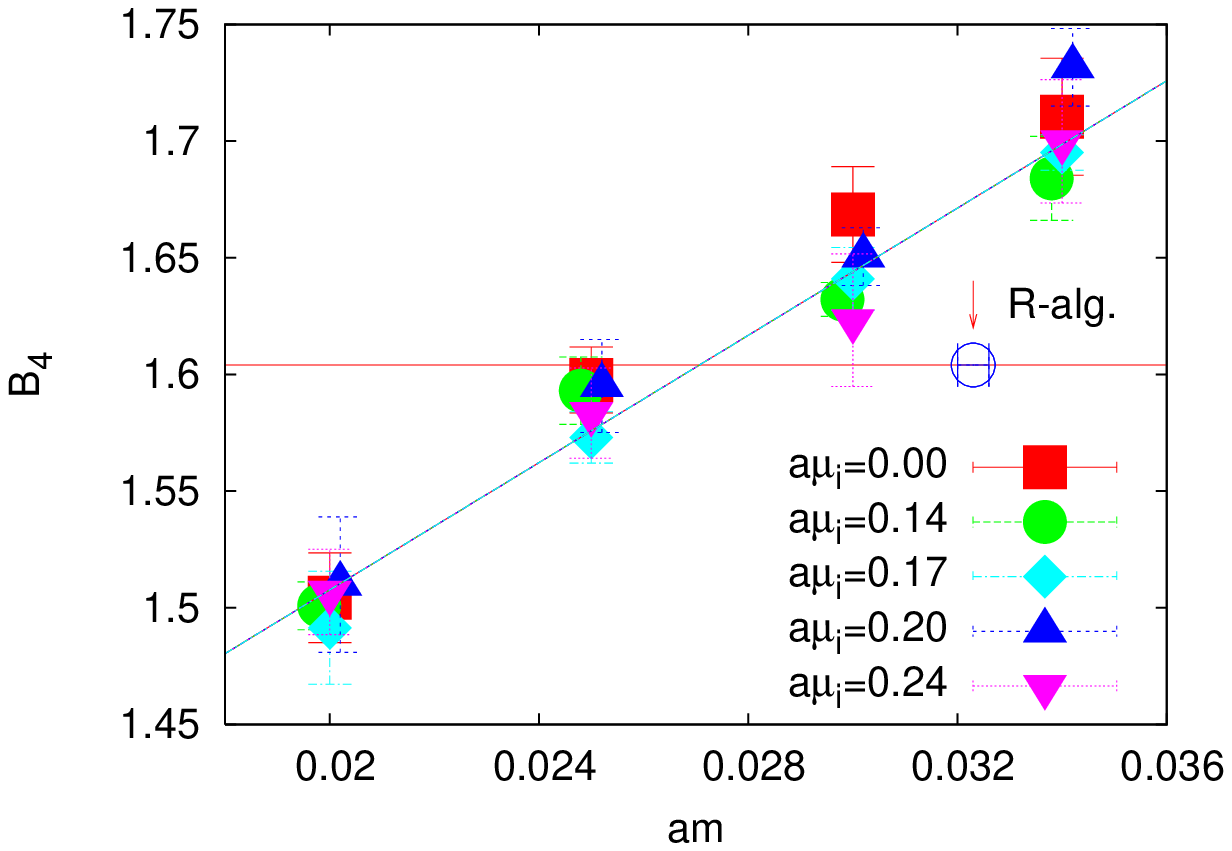}}
\scalebox{0.62}{\includegraphics{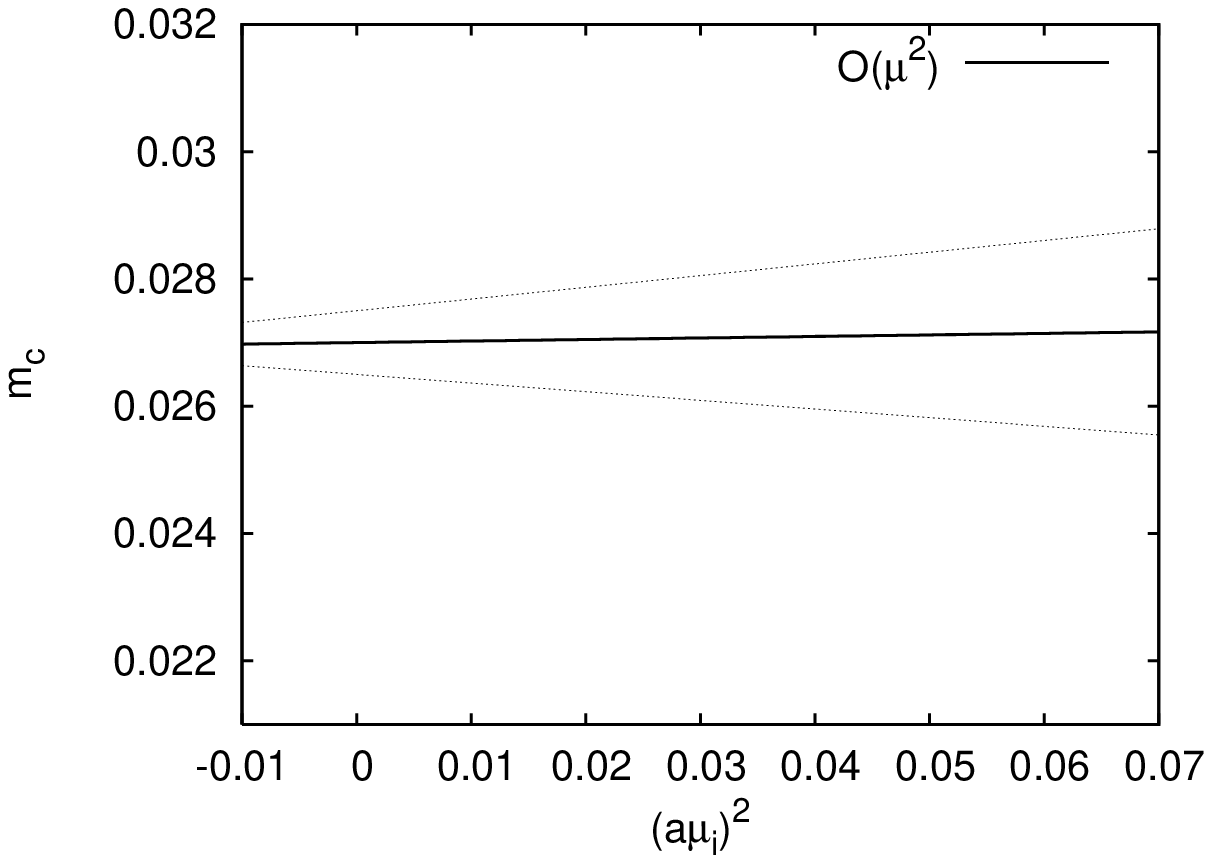}}
}
\caption{Left: $B_4(am,a\mu_I)$ for different imaginary chemical potentials.
Right: One-sigma error band for the critical mass $a m^c(a \mu_I)$ resulting
from a linear fit.}
\end{figure}

We now turn on an imaginary chemical potential $\mu=i\mu_I$, and for
each $\mu_I$ monitor the Binder cumulant $B_4$ as a function of the
quark mass. Our results are summarized in Fig.~3, left. The chemical
potential has almost no influence on $B_4$. 
A lowest-order fit, linear in $a m$ and $(a \mu)^2$, gives the error band Fig.~3, right, 
corresponding to
\be
a m^c(a \mu) = 0.0270(5) - 0.0024(160) (a \mu)^2
\label{c_prime1}
\ee

Care must be taken for the conversion to physical units. The crucial
point is that, as we increase the chemical potential $\mu_I$, we tune
the gauge coupling $\beta$ upwards to maintain criticality, so that $a(\beta)$
decreases: our observation that $a m^c(\mu_I) \approx const.$ does {\em not}
mean that $m^c(\mu_I) \approx const.$, but that $m^c(\mu_I)$ increases
with $\mu_I$, or {\em decreases} with a real chemical potential $\mu$.
If we express 
\bea
\label{c1prime}
\frac{a m^c(\mu)}{a m^c(0)} = 1 + \frac{c'_1}{a m^c(0)} (a \mu)^2 + ... \\
\frac{m^c(\mu)}{m^c(0)} = 1 + c_1 \left( \frac{\mu}{\pi T} \right)^2 + ...
\label{c1}
\eea
then $c_1$ and $c'_1$ are related by
\be
c_1 = \frac{\pi^2}{N_t^2} \frac{c'_1}{a m^c(0)} + \left( \frac{1}{T_c(m,\mu)} \frac{d T_c(m,\mu)}{d(\mu/\pi T)^2} \right)_{\mu=0}
\ee
where $m = m^c(\mu)$ in the second term. Writing the transition temperature as
\be
\frac{T_c(m,\mu)}{T_c(m^c_0,0)} = 1 + A \frac{m - m^c_0}{\pi T} + B \left( \frac{\mu}{\pi T} \right)^2 + ...
\ee
one obtains
\be
c_1 = \left( B + \frac{\pi^2}{N_t^2} \frac{c'_1}{a m^c(0)} \right) \left( 1 - A \frac{m^c_0}{\pi T} \right)^{-1} \quad .
\ee
$c'_1$ and $\frac{m^c_0}{\pi T}$ are both small, so that $c_1$ is nearly 
equal to $B$. Estimates of $B$ and $A$ can be obtained by converting our result for the
pseudo-critical gauge coupling
\be
\beta_0(am, a\mu) = 5.1369(3) + 1.94(3) (a m - a m^c_0) + 0.781(7) (a \mu)^2 
\ee
to physical units. Using the 2-loop $\beta$-function gives $A=2.111(17), B=-0.667(6)$
so that finally
\be
\frac{m^c(\mu)}{m^c(0)} = 1 - 0.7(4) \left( \frac{\mu}{\pi T} \right)^2 + ...
\label{c1phys}
\ee
The error above is conservative and includes the uncertainty from using
different fitting forms (see Table 2, Ref.~\cite{OP_last}).
The main source of systematic error comes from using the 2-loop $\beta$-function to
obtain $B$. 
The non-perturbative $\beta$-function varies more steeply and may increase
$A$ and $B$, in magnitude, by up to a factor 2. This will make $c_1$ more
negative.

We thus have clear evidence that, in the $N_f=3$ theory on an $N_t=4$ lattice,
the region of first-order transitions {\em shrinks} as a baryon chemical potential
is turned on, and the ``exotic scenario'' of Fig.~2, right, is the correct one.
This result is further supported by recent simulations of the same theory, under
an isospin chemical potential~\cite{DKS_mc}.

\section{$N_f=2+1$}
\label{sec_QCD}

\begin{figure}[t!]
\centerline{
\includegraphics[width=8.0cm]{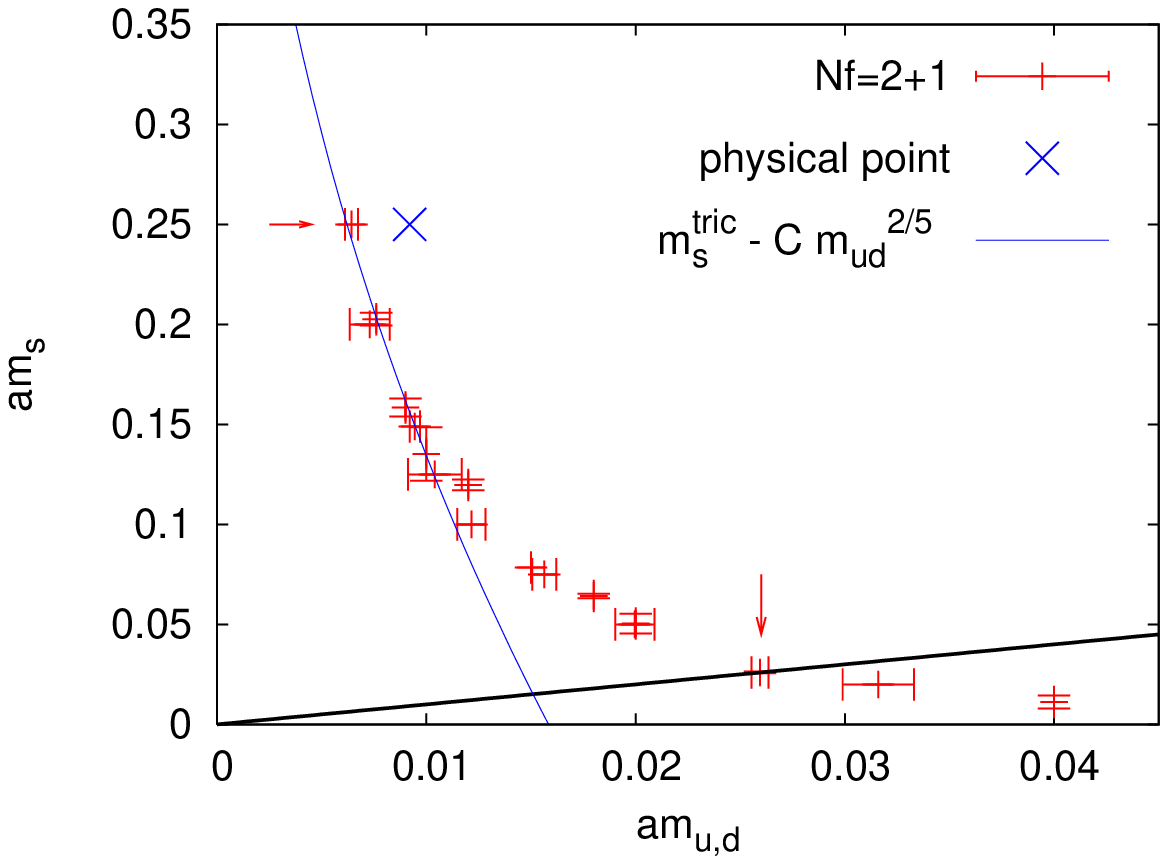}
\includegraphics[width=8.0cm]{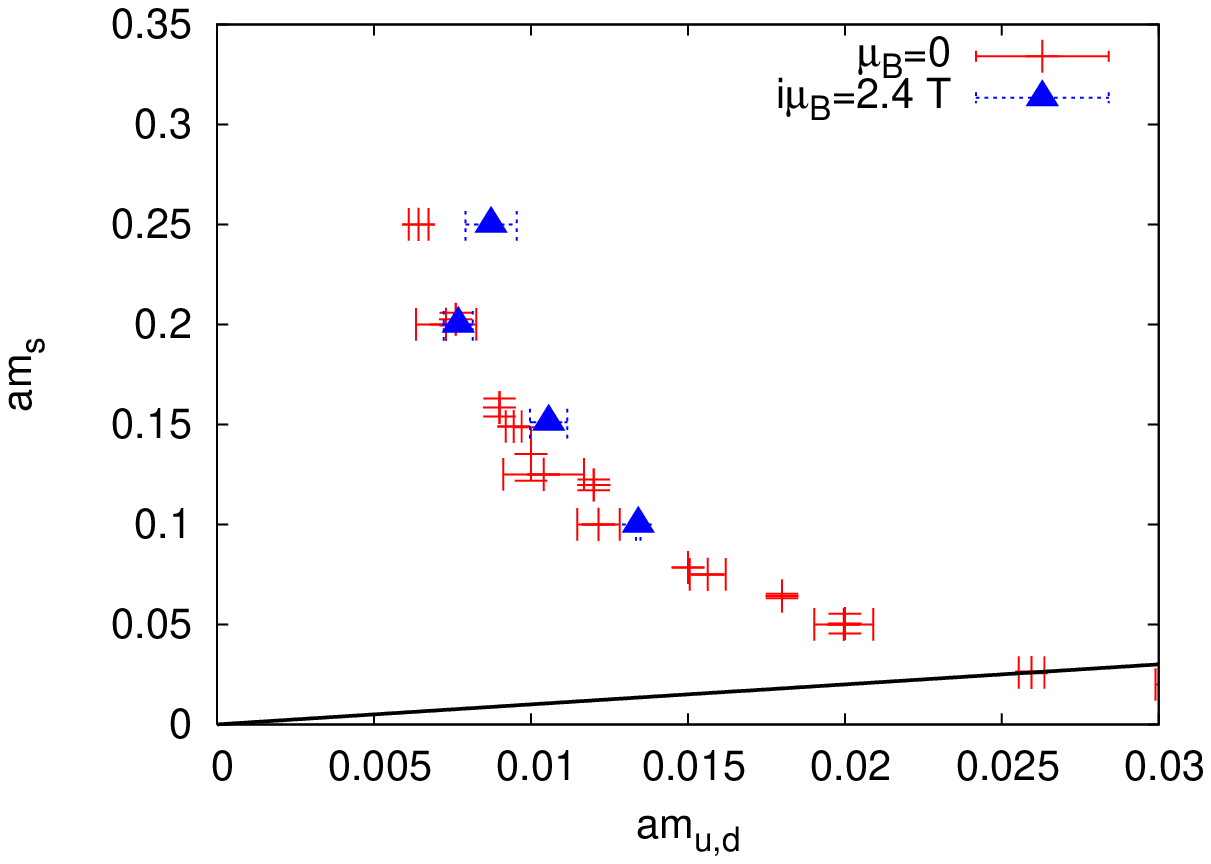}
}
\caption{Left: The chiral critical line in the bare quark mass plane at $\mu=0$.
$N_f=3$ is shown by the solid line. Also shown are the physical point according to
\cite{FK_phys}, and a fit corresponding to a tricritical point $m_s^{tric}\sim 2.8 T$.
Right: Comparison of the critical line at $\mu=0$ and $a \mu_I = 0.2$.}
\end{figure}

We now proceed to the non-degenerate case. First, at $\mu=0$, we map out the line
of second-order transitions in the $(a m_{u,d},a m_s)$ plane. 
Our results, shown Fig.~4, left, are in qualitative agreement with expectations Fig.~1.
In particular, they are consistent with the possible existence of a tricritical point
$(m_{u,d}=0,m_s=m_s^{tric})$. Using its known, Gaussian exponents, our data favor
(blue line in Fig.~4 left) a heavy $m_s^{tric} \sim 2.8 T_c$.

A more immediate issue is whether the QCD physical point lies on the crossover side
of the critical line as expected. For that purpose, we have performed spectrum 
calculations at $T\sim 0$, at the parameters corresponding to the horizontal arrow
in Fig.~4 left ($a m_{u,d}=0.005, a m_s = 0.25, \beta=5.1857$). They show that
$m_s$ is approximately tuned to its physical value 
($\frac{m_K}{m_\rho} \sim \frac{m_K}{m_\rho}|_{\rm phys}$),
while the pion is lighter than in QCD ($\frac{m_\pi}{m_\rho}=0.148(2) < 0.18$).
This confirms that the physical point lies on the right of the critical line,
i.e. in the crossover region 
\footnote{In fact, our estimate of the lattice parameters
corresponding to the physical point is consistent with that of Fodor \& Katz
using the same action, but the R algorithm~\cite{FK_phys}.}.
This conclusion has been confirmed by very recent calculations on finer 
lattices~\cite{FK_Nature}.
Also, we find $T_c$ to vary little along the critical line, 
in accordance with model calculations~\cite{Szep}.

We now couple an imaginary chemical potential $a \mu_I = 0.2$ to the two light flavors,
and measure the change in the critical mass $a m_{u,d}$ as in the $N_f=3$ case.
Fig.~4 right shows the same trend as for $N_f=3$: the critical mass is constant
or slightly increasing, {\em in lattice units}. The conversion to \linebreak physical units
proceeds as in eqs.(\ref{c1prime}-\ref{c1phys}). 
Since the critical gauge coupling $\beta_0(a \mu_I)$ increases
with $\mu_I$, the coefficient $B$, which is the dominant contribution to $c_1$,
is negative. Together with a very small or slightly negative value for $c'_1$, 
it implies again that the first-order region {\em shrinks} as the baryon chemical
potential is turned on, and the ``exotic scenario'' of Fig.~2, right, is the correct one.

This statement comes with several caveats: $(i)$ our lattice is very coarse
($a \sim 0.3$ fm); $(ii)$ as we consider lighter $m_{u,d}$, our box becomes small 
($m_\pi L \sim 1.7$ for the worst case);
$(iii)$ we use ``rooting'' of the staggered determinant to simulate 1 and 2 flavors,
albeit our measure is positive with an imaginary $\mu$, so that we avoid the pitfalls
of \cite{root}.

\section{Discussion}

Our results appear in qualitative contradiction with those of Fodor \& Katz~\cite{FK_phys}
and of Gavai \& Gupta~\cite{GG}, which both conclude for the existence of a critical point
$(\mu_E,T_E)$ at small chemical potential $\mu_E/T_E \lesssim 1$. 
Let us consider the reasons for such disagreement.

\begin{figure}[h!]
\centerline{
\includegraphics[width=8.0cm]{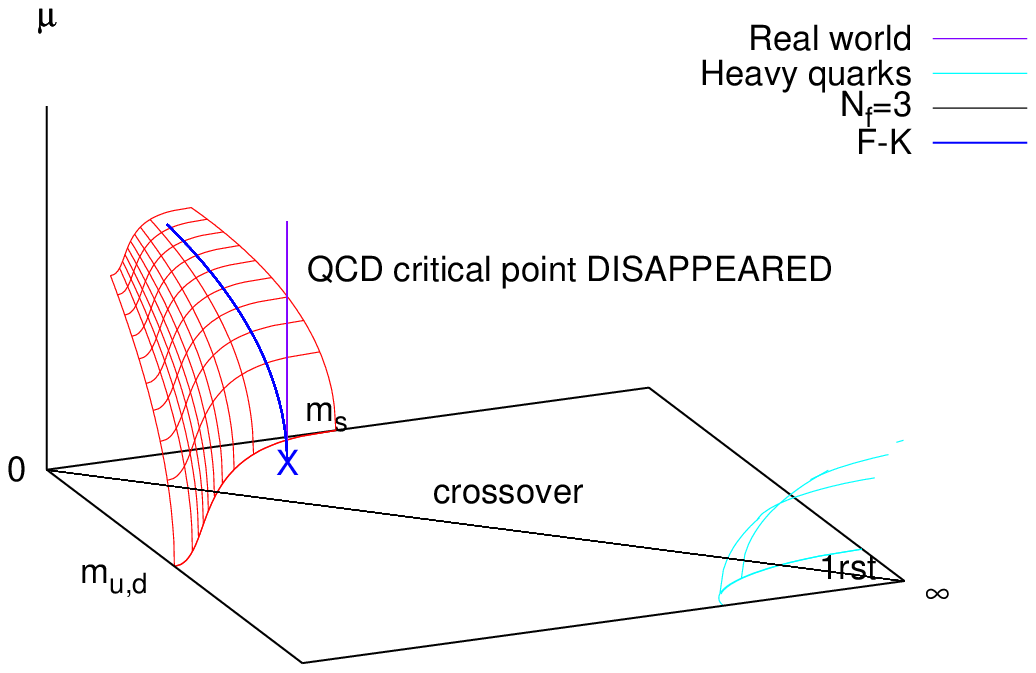}
\includegraphics[width=8.0cm]{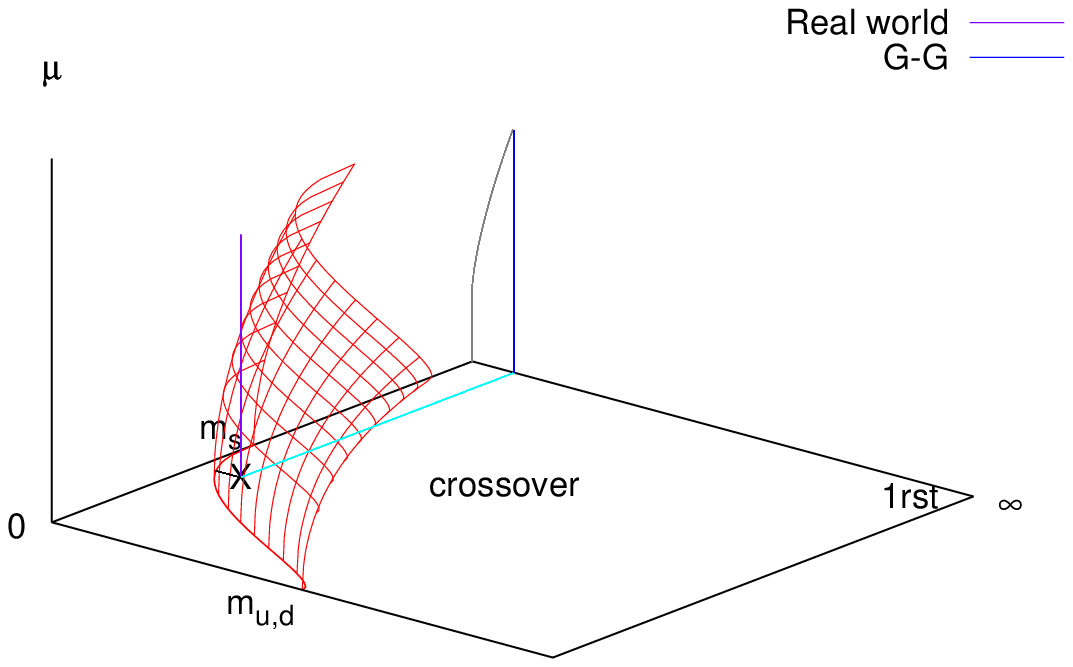}
}
\caption{Left: Effect of keeping the quark mass fixed in lattice units in \cite{FK_phys}.
Right: Comparison at finite $\mu$ between the $N_f=2+1$ and the $N_f=2$ theory
considered in \cite{GG}.}
\end{figure}

$\bullet$ Fodor \& Katz obtain Monte Carlo results at $\beta=\beta_c, \mu=0$, and perform
a double reweighting in $(\beta,\mu)$ along the pseudo-critical line $\beta_c(a\mu)$.
By construction, this reweighting is performed at a quark mass fixed {\em in lattice units}:
$a m_{u,d}=\frac{m_{u,d}}{T_c} = const.$. Since the critical temperature $T_c$ decreases 
as they turn on $\mu$, so does their quark mass. This decrease of the quark mass pushes
the transition towards first order, which might be the reason why they find a critical
point at small $\mu$. This effect is illustrated in the sketch Fig.~5, left, where the bent
trajectory \`a la Fodor \& Katz intersects the critical surface, while the vertical line
of constant physics does not.

Put another way, Fodor \& Katz measure the analogue of eq.(\ref{c1prime}) instead of (\ref{c1}).
From their Fig.~1 (Ref.~\cite{FK_phys}), the coefficient $c'_1$ which one would extract
would be essentially zero like ours. As in our case, the variation of $T_c$ with $\mu$
makes a dominant contribution, which may change the results qualitatively.

$\bullet$ Gavai \& Gupta try to infer the location of the critical point by estimating
the radius of convergence of the Taylor expansion of the free energy in $(\mu/\pi T)^2$.
Regardless of the systematic error attached to such estimate when only 4 Taylor coefficients
are available, we want to point out that they consider a theory without strange quark,
i.e. $N_f=2$ only. The $(\mu,T)$ phase diagram of such a theory is qualitatively different
from that of $N_f=2+1$ QCD. At $\mu=0$, the order of the finite-temperature transition
as $m_{u,d} \to 0$ is not settled~\cite{DiG}. Assuming a second-order $O(4)$ 
transition, one expects then a tricritical point at $(m_{u,d}=0, \mu=\mu^{tric})$,
beyond which a non-zero critical mass $m_{u,d}^c(\mu)$ can be defined, as sketched
in Fig.~5 right. The quantitative relevance of results, even accurate, for this
$N_f=2$ theory to QCD is unclear to us.

Therefore, we find no inconsistency between our results and those above.
We conclude that the existence of a critical point $(\mu_E,T_E)$ in QCD at
small chemical potential $\mu_E/T_E \lesssim 1$ is an open question.
Our numerical evidence, with the caveats mentioned in Sec.~\ref{sec_QCD},
is that the curvature of the critical surface is as illustrated Fig.~2 right.
Our main systematic error comes from our coarse lattice spacing $a \sim 0.3$ fm~\cite{Karsch}.
If confirmed on a finer lattice, the implications of our finding are as follows.
In the region where a leading Taylor expansion of the critical surface is a good 
approximation, i.e. $\mu/T \lesssim 1$, corresponding to the experimentally
accessible regime, no critical point exists which is analytically connected to $\mu=0$.
Of course, we cannot exclude that the QCD phase diagram is more complex, 
and partly inaccessible to our imaginary $\mu$ + Taylor expansion strategy.

\end{document}